\documentstyle[12pt,aaspp4]{article}
\received{1997 March 20}

\def\etal{{et al.}}
\def\asca{{\it ASCA}}

\def\pcmsq{{$\rm cm^{-2}$}}

\def\nh{{$N_{\rm H}$}}

\newbox\grsign \setbox\grsign=\hbox{$>$} 
\newdimen\grdimen \grdimen=\ht\grsign
\newbox\simlessbox \newbox\simgreatbox \newbox\simpropbox
\setbox\simgreatbox=\hbox{\raise.5ex\hbox{$>$}\llap
     {\lower.5ex\hbox{$\sim$}}}\ht1=\grdimen\dp1=0pt
\setbox\simlessbox=\hbox{\raise.5ex\hbox{$<$}\llap
     {\lower.5ex\hbox{$\sim$}}}\ht2=\grdimen\dp2=0pt
\setbox\simpropbox=\hbox{\raise.5ex\hbox{$\propto$}\llap
     {\lower.5ex\hbox{$\sim$}}}\ht2=\grdimen\dp2=0pt
\def\simgreat{\mathrel{\copy\simgreatbox}}

\begin{document}

\title{ASCA observations of type-2 Seyfert Galaxies: 
II The Importance of X-ray Scattering and Reflection}

\author {T.J.Turner \altaffilmark{1,2}, 
I.M. George \altaffilmark{1, 2},
K. Nandra \altaffilmark{1, 3},  
R.F. Mushotzky\altaffilmark{1}, \nl
 }

\altaffiltext{1}{Laboratory for High Energy Astrophysics, Code 660,
	NASA/Goddard Space Flight Center,
  	Greenbelt, MD 20771}
\altaffiltext{2}{Universities Space Research Association}
\altaffiltext{3}{NAS/NRC Research Associate}

\slugcomment{Submitted to {\em The Astrophysical Journal}}

\begin{abstract}

We discuss the importance of X-ray scattering and Compton reflection
in type-2 Seyfert Galaxies, based upon the analysis of {\it ASCA} 
observations of 25 such sources.   
Consideration of the iron K$\alpha$, \verb+[+O{\sc iii}\verb+]+ line and X-ray variability 
suggest that NGC~1068, NGC~4945, NGC~2992, Mrk~3, Mrk~463E and Mrk~273 
are dominated by reprocessed X-rays.
We examine the properties of these sources in more detail.

We find that the iron K$\alpha$ complex contains significant 
contributions from neutral and high-ionization species of iron.  
Compton reflection, hot gas and 
starburst emission all appear to make significant contributions to the 
observed X-ray spectra. 

Mrk~3 is the only source in this subsample which does not have a 
significant starburst contamination. The {\it ASCA} spectrum 
below 3 keV is dominated by hot scattering gas 
with $U_{X}\sim 5$, $N_{H} \sim 4 \times 10^{23} {\rm cm}^{-2}$.  This 
material is more highly ionized than the zone of material comprising
the warm absorber seen in Seyfert~1 galaxies, but may 
contain a contribution from shock-heated gas associated with the jet.  
Estimates of the X-ray scattering fraction cover 0.25 -- 5\%. The spectrum
above 3 keV appears to be dominated by a Compton reflection component 
although there is 
evidence that the primary continuum component becomes visible close to 
$\sim10$ keV.

\end{abstract}


\section{Introduction}

\label{sec:intro}

The optical spectra of most nearby active galactic nuclei (AGN) 
are dominated by narrow optical 
lines (e.g. Lawrence 1991) and are therefore broadly classified as type-2 
Seyfert galaxies. However, broad optical lines 
with characteristics similar to 
those observed in Seyfert~1 galaxies were discovered in the polarized
light from NGC~1068 (Antonucci \& Miller 1985).  It was suggested that
the polarization arose as a result of electron scattering, implying
that the optical broad-line-region in NGC~1068 is only visible in
scattered light. When several more examples of the phenomenon were
found (Miller \& Goodrich 1990) a general model emerged in which the
nuclear spectra of both Seyfert types 1 and 2 are the same, but the
nuclear regions of type~1 are observed directly while Seyfert 2 nuclei
are hidden behind a large column of obscuring material.  The simplest
geometry for the obscuring material is a torus. This allows
orientation-dependant obscuration, and hence can explain the existence
of the type-1 and type-2 Seyfert galaxies. Hubble Space Telescope ({\it HST})
observations have now resolved a torus in the AGN NGC~4261
(Ferrarese, Ford \& Jaffe 1996).  Such a torus allows the 
existence of 
some lines-of-sight for which radiation from within the torus can only
be viewed via scattering.  The high opacity of the obscuring material 
makes it 
difficult to directly observe the central engines of Seyfert~2
galaxies at most wavelengths.  Medium energy X-rays can penetrate
column densities of up to $\sim 10^{24}$~\pcmsq, allowing a direct
view of some type-2 nuclei, but observations of sources with even
larger columns should 
reveal predominantly scattered or reflected X-rays below 10 keV.

In a companion paper (Turner \etal\ 1997, hereafter Paper I) we
presented the data analysis results of {\it ASCA} observations of a
sample of 17 Seyfert 2 galaxies, plus 8 Narrow Emission Line Galaxies
(NELGs).  Paper I dealt with temporal and spectral X-ray results from
those two classes of AGN together.  The observations were drawn from
the public archive, and the sources do not comprise a complete sample.
The data selection, reduction and analysis are described in detail in
Nandra \etal\ 1997 (hereafter N97) and Paper~I.  The sources in which 
we observe the nucleus through the absorber will be discussed in  
another paper (Turner \etal\ 1997b). In the present paper 
we determine which sources from the original sample are most likely to
be dominated by reprocessed X-ray emission, and then examine the
properties of this subset in detail. 

\section{Signatures of Reprocessed X-ray Emission}

Iron K$\alpha$ lines in AGN arise via reprocessing of the primary
X-rays.  If K$\alpha$ lines of very high equivalent width are observed, this
is an indication that one is viewing a spectrum dominated by
reprocessing. For Seyfert 2 galaxies, it is appropriate to consider
two types of reprocessing, which we shall refer to as 
``Compton reflection'' and ``scattering''. We use the term Compton
reflection (or simply ``reflection'') to mean reprocessing of X-rays via
Compton scattering and fluorescence by material which
is optically thick to electron scattering.  We use the term
``scattering'' to apply to the case where the optical depth to 
electron scattering is $ 0 < \tau < 1$.

In Seyfert 1 galaxies, the iron K$\alpha$ lines are thought to arise
via Compton reflection from an accretion disk (e.g., N97), which
produces lines of equivalent width $\sim$~few hundred eV with respect
to the directly observed continuum (e.g. George \& Fabian 1991).
However, the equivalent width of the iron K$\alpha$ line can be as
high as a few keV with respect to the Compton-reflected continuum
(George \& Fabian 1991; Matt \etal\ 1991). 
Such a situation is somewhat unlikely in the case of an
accretion disk, as the central continuum and iron line are both
produced in a very small region and obscuring one without the other
would require an unusual geometry. However, the reflection may also
occur at the inner surface of the obscuring material, possibly the 
torus, and the orientation and 
opening angle can then allow a view of the inner surface, but not the
illuminating source.  Some early {\it ASCA} results showed that
Compton reflection of nuclear radiation from optically thick material
can be an important component of Seyfert 2 galaxies. The Circinus
galaxy (Matt \etal\ 1996) and NGC 6552 (Fukazawa \etal\ 1994) show
strong reflection components dominating the spectra even below 10 keV.

If the iron K$\alpha$ line is produced in optically thin 
gas then its observed equivalent width can also be 
large as long as the direct continuum is not observed.
The line equivalent width is a strong function of the ionization-state
of the material (which is also the case for the reflected iron lines,
but both the disk and torus are thought to be relatively neutral).
For low ionization-states the equivalent width of the iron K-shell
line, measured against the scattered nuclear emission (plus free-free
and bound-free emission from the scattering gas) can be as high as
$\sim 1$ keV for the 6.4 keV line (Netzer 1996), and could 
appear to be up to a few keV for an iron line blend. 

Although it is tempting to interpret lines of high equivalent width as
evidence for a dominant reflected or scattered component, high
equivalent widths can also arise if the ionizing radiation is 
anisotropic (e.g. Ghisellini \etal\  1991) or 
if there is a significant lag between
a fall in the continuum flux and the reprocessed spectrum.  Indeed,
Weaver \etal\ (1996) suggested a 16 year lag between a drop in
continuum and line fluxes, to explain the large equivalent width in
the iron K$\alpha$ line for NGC~2992 at the {\it ASCA} epoch. 
Thus, when assessing whether
or not a source is dominated by reprocessing, it is important to
take other evidence into account.

An alternative method of determining whether a type-2 Seyfert
galaxy  contains a
hidden X-ray source is by examining the flux in the \verb+[+O{\sc iii}\verb+]+ 
$\lambda 5007$ emission line 
(hereafter simply referred to as \verb+[+O{\sc iii}\verb+]+) 
observed in the optical regime.  Mulchaey \etal\ (1994) 
found a strong 
correlation between \verb+[+O{\sc iii}\verb+]+  and hard X-ray flux in a sample of
Seyfert~1 galaxies.  This provides an independant
 estimate of the nuclear X-ray
luminosity for Seyfert 2 galaxies for comparison with
the observed X-ray luminosity. If the observed X-ray luminosity is
substantially less than that predicted by \verb+[+O{\sc iii}\verb+]+  this is an
indication that the X-ray emission is being observed indirectly, with
only a fraction of the total luminosity reaching our line-of-sight.

\section{Identification of the reflected/scattered sources} 

A summary of the {\it ASCA} sample is detailed in Table~1. We now
consider the properties of these sources in relation to the above
discussion, to determine which Seyfert 2 galaxies are likely
to be dominated by reprocessed emission.

\subsection{Iron line equivalent width}

In Paper~I we found the 6-7 keV regime of low redshift type-2 
Seyfert galaxies  to
be dominated by emission at 6.4 keV. This indicates that a large
fraction of the iron K$\alpha$ line flux is produced in gas with
ionization-state $<$ Fe {\sc XVI}.  Furthermore, such a line is not
generally observed in starburst galaxies (Ptak \etal\ 1997), and
certainly not with such a high equivalent-width, indicating that these
lines are related to the active nucleus.  We found such a line in
$72$\% of the sample sources at $> 99\%$ confidence.  Fig.~1
(adapted from a figure in Paper~I) shows the equivalent widths of the
6.4 keV lines versus X-ray absorbing column (except in the case of
NGC~6251, where the 6.68 keV line is the only iron K$\alpha$ line
detected, and so this is plotted).  All detected sources are shown
from Paper~I, except NGC~1667 which had a best-fit line equivalent
width of zero, and an upper limit of 3 keV.  The line equivalent
widths are measured against the hard X-ray continuum component
(i.e. the component dominating the fit over 6 - 8 keV, from the
best-fit continuum model, see Paper~I for details).
First we consider two plausible processes which could produce the line
emission {\it without} invoking a dominant reflected or scattered
component. 

Line emission could be produced by transmission through the
line-of-sight absorber.   
The dot-dash line in Fig~1 denotes the equivalent width of iron
K$\alpha$ predicted to be produced by a uniform shell of neutral
material with solar abundances subtending 4$\pi$ to a continuum source 
of photon index $\Gamma=2.0$ (Leahy \& Creighton
1993).  It is immediately obvious that a substantial fraction of the emission
lines in our sample cannot be produced by such a process.
Non-uniformity of the absorbing material cannot 
explain these discrepancies. If we view the source with a line-of-sight 
through absorption lower than the mean 
column, then we see more continuum photons at 6.4 keV, reducing the 
apparent equivalent width of the iron K$\alpha$ line. In the
opposite case, enhancements in the line equivalent width can be
produced, but large equivalent widths ($\simgreat 100$~eV) are only
produced when the line-of-sight column is very high $\gg
10^{23}$~\pcmsq. This is not the case for the sources which are
discrepant with the Leahy \& Creighton line.

The mean equivalent width of the emission lines in Seyfert 1 galaxies
is $230 \pm 60$~eV (N97), and most of this is thought to be 
attributable to Compton reflection of nuclear X-rays from the 
surface of the putative accretion disk. 
Further enhancement of this value in Seyfert 
2 galaxies could be afforded if the direct continuum is suppressed by
absorption, but the Compton-reflection component is not. As we have
noted above, if the Compton reflection arises in an accretion disk,
this is not very likely. It is, however, plausible if the 
absorbing material (possibly, but not necessarily the torus) is
responsible for the Compton reflection. The dashed line (Fig.~1) shows the
predicted equivalent width from reflection as a function of \nh,
assuming that only the power-law is absorbed, but that the reflection
component remains unchanged.   Many of our
sources lie significantly above this line also, which in any case only
implies large ($\sim 1$~keV) equivalent widths for very high columns
($\gg 10^{24}$~\pcmsq).  The direct continuum in the 2-10 keV band is
then so heavily suppressed that the sources are dominated by the
reflection component anyway. 

In cases where a source lies significantly above both of these two
model lines, we consider there to be good evidence that the \asca\
spectrum is dominated by reprocessed emission. These are NGC~1068;
Mrk~3; NGC~2992 and NGC~4945.  A number of other sources (Mrk~463E;
NGC~6251; NGC~4968; NGC 5135 and Mrk 273) have best-fit equivalent
widths higher than the model lines, but are formally consistent with
one or both. These sources are could plausibly be dominated by
reprocessing, but such a conclusion is not warranted based on the
K$\alpha$ line strengths alone and require supporting evidence.  We
now consider such evidence.

\subsection{O{\sc iii} measurements}

Table 1 shows the 2-10 keV absorption-corrected luminosity
\footnote{H$_0=50\ {\rm km\ s}^{-1} {\rm Mpc}^{-1}$ and q$_0=0.5$ assumed 
throughout}
predicted based upon \verb+[+O{\sc iii}\verb+]+  line 
flux measurements (see Mulchaey 
\etal\ 1994, Polletta \etal\ 1996 and references therein), and the
measured (absorption-corrected) 2-10 keV luminosity for 
each source.  Mulchaey \etal\ (1994) find the relationship between 
hard X-ray flux and \verb+[+O{\sc iii}\verb+]+   flux is 
consistent with being the same for the Seyfert~1 and Seyfert~2 galaxies in 
their sample.  There 
is a larger scatter in the distribution in the 
case of the Seyfert~2 sample, as might be 
expected because that sample contains some sources for which 
the hard X-ray flux is just the scattered X-rays. For that reason, we 
use the more tightly-determined correlation for the Seyfert~1 galaxies as the 
basis of our calculation. The resulting intrinsic luminosities are 
tabulated and plotted against the observed
values (from Paper~I, Table~12) in Fig.~2.

Unfortunately the \verb+[+O{\sc iii}\verb+]+  flux can include some unresolved
contribution from starburst emission, depending on the aperture used
and the level and location of starburst activity in the host.  We
might expect some predicted luminosities to be too high because of
this confusion. We also expect scatter in the plot due to variability
and the imperfect nature of the correlation between \verb+[+O{\sc iii}\verb+]+ 
 and 
hard X-ray luminosity.  Given these uncertainties we highlight (by
naming the sources on the plot) only those sources whose predicted
luminosities exceed the observed by a factor $>15$. This confirms the
suggestion of a hidden X-ray continuum in NGC~1068, Mrk~3 and NGC~2992,
which we had already concluded based on the iron line evidence.
The lack of an \verb+[+O{\sc iii}\verb+]+  flux measurement for NGC~4945
excludes that source from this test.  However, heavy obscuration in
NGC~4945 is well established.  Done \etal\ (1996) used the combined
{\it Ginga}, {\it ASCA} and {\it OSSE} data to confirm the Iwasawa
(1994) result that the nucleus of NGC~4945 is so heavily absorbed that
the direct component is only visible above 10 keV. 

Examination of the \verb+[+O{\sc iii}\verb+]+  flux also indicates a hidden X-ray
continuum in Mrk~273, Mrk 463E, NGC~4507, NGC~1667 and NGC~5695.  The
cases for reprocessing in Mrk~273 and Mrk~463E now become much
stronger, as the iron K$\alpha$ lines also had a very high equivalent
width (although the constraints were such that an alternative origin
could not be ruled out).  The iron K$\alpha$ equivalent widths
indicate consistency with an origin in the line-of-sight absorber in
NGC~4507 and NGC~1667 (although the latter in particular is highly
uncertain).  Thus we conclude that the evidence for extensive
reprocessing in these cases is inconclusive and we do not consider
NGC~4507 and NGC~1667 in detail in this paper.  NGC 5695 has no
observed {\it ASCA} flux (just an upper limit) and so it is not
plotted in Fig.~2.  However, if the intrinsic luminosity implied by
the \verb+[+O{\sc iii}\verb+]+ flux was observed directly, NGC~5695 
would have been 
detected easily by \asca. This suggests that the absorbing column
exceeds a few $10^{24} {\rm cm}^{-2}$.
Elongated narrow-line regions in some Seyfert galaxies appear to be ionized by
a more intense radiation field than is directly observable, (e.g. Haniff,
Wilson \& Ward 1988, Pogge 1988, Tadhunter and Tsvetanov 1989). This provides
supporting evidence for hidden nuclei in sources such as Mrk~3.

Finally, we note that NGC~6251 and NGC~4968, which showed (albeit
poorly-constrained) iron K$\alpha$ lines with high equivalent width do
not show evidence for a hidden continuum based on 
the \verb+[+O{\sc iii}\verb+]+ 
measurements, and so we do not consider them further in this paper.
 
\subsection{X-ray variability}

Of the sources most likely to be dominated by reprocessed emission,
only NGC~1068 was bright enough for us to search for rapid X-ray
variability (Paper I). It did not vary significantly when sampled on a
128 s timescale.  While several of the sources appear to show hard
X-ray variability on timescales of years (Polletta \etal\ 1996), all
cases except Mrk~3 are based upon comparison between {\it HEAO/A1} and
{\it ASCA} or {\it Ginga}. The low angular resolution of {\it HEAO/A1}
yields systematically high fluxes assigned to those measurements of
many source fluxes in the Polletta \etal\ (1996) catalog. Thus we do
not consider apparent flux variations which depend on the {\it
HEAO/A1} measurements, to be reliable.

NGC 2992 is unique amongst the reprocessing-dominated sources because
it shows correlated variability in the hard and soft X-ray regimes
down to timescales of a few days (Weaver \etal\ 1996, Turner \& Pounds
1989). This indicates that we are seeing the 0.5 - 10 keV flux
directly.  Thus Weaver \etal\ (1996) 
suggest that NGC~2992 has a strong observed iron K$\alpha$ line
because of a lag between direct and reprocessed flux, rather than due
to obscuration of the nucleus. However, the 6.4 keV line is still
inferred to arise by Compton reflection, and so we continue to include
it in our consideration of the reprocessed source spectra.

Thus, taking the iron K$\alpha$, \verb+[+O{\sc iii}\verb+]+ line and variability
evidence together, we are most confident that the {\it ASCA}
observations are dominated by reprocessed X-rays in the cases of
NGC~1068, NGC~4945, NGC~2992, Mrk~3, Mrk~463E and Mrk~273.  
For ease of reference we now denote these as ``group C'' sources 
and these are  marked with squares on Fig.~1. 
For later reference we denote the sources lying on the 
Leahy and Creighton line to be 
``group A'', and those with iron K$\alpha$ equivalent widths lying between 
that line and 230 eV to be ``group B''. Group B sources have iron K$\alpha$ 
lines consistent with those observed from the relativistic accretion 
disk in Seyfert~1 galaxies (N97). The properties of group A and B 
sources will be discussed in a later paper (Turner \etal\ 1997b).  We 
continue here by investigating what the X-ray properties of 
group C sources can tell us about conditions in the reprocessing media.

\section{The nature of the reprocessing regions}

\subsection{Constraints from absorption}

The absorbing columns detected in the reprocessed sources ($\sim
10^{22} {\rm cm}^{-2}$) do not necessarily 
represent the line-of-sight attenuation
to the nucleus.  Rather, they may represent opacity in the scattering
gas.  In principle, then, measurements of that absorption could be
used to indicate the column density and ionization state of that
medium.  However, misleading spectral fit parameters can be obtained
when fitting simple continuum models (Paper~I) to spectra dominated by
photoionized emission.  To illustrate this point, Fig.~3 shows a
typical emission spectrum from ionized absorbing gas observed in the
line-of-sight to $\gtrsim 60\%$ of Seyfert~1 galaxies (Reynolds 1997;
George \etal\ 1997). The gas has an ionization parameter $U_X \sim
0.125$ and column $N_H \sim 4 \times 10^{22} {\rm cm}^{-2}$, based on
the analysis of {\it ASCA} observations of Seyfert~1 galaxies (see
George \etal\ 1997 for details and the definition of $U_X$).  For a
demonstration we simulated a 40 ksec {\it ASCA} observation of the
emission spectrum of the warm absorber gas and then fit the simulated
spectrum with a model composed of an absorbed power-law plus a
Raymond-Smith equilibrium plasma. We find a best-fitting spectrum with
a flat power-law of $\Gamma=0.87$ absorbed by a column 
$N_H = 1.8 \times 10^{22} {\rm cm}^{-2}$ plus an 
unabsorbed plasma with $kT=0.86$
keV, with a fit-statistic of $\chi^2=424/331$.  Both the fit
parameters and acceptability are reminiscent of values obtained when
fitting some Seyfert~2 galaxies with simple models
(e.g. Paper~I). Apparently, the presence of the emission spectrum from
the scattering gas can mimic an absorbed flat power-law plus a soft
thermal component.

If the warm absorber commonly seen in Seyfert~1 galaxies is the same
material which scatters nuclear X-rays into our line of sight in the
Seyfert~2 case, then we should expect the same distribution of
ionization parameter and column for these components, and we discuss
this in \S6, in the context of the spectral results from Mrk~3. Krolik
\& Kriss (1995) suggest that an X-ray heated wind can produce both the
warm absorber in Seyfert~1 galaxies, and warm scattering gas affording
a view of the central nuclei of Seyfert 2 galaxies. The wind idea is
supported by the blueshifts observed in the UV absorption systems
in some QSOs and Seyfert galaxies. Those systems generally show
blueshifts indicative of outflow velocities of a few hundred
km/s. Observations to date of the X-ray absorber have been insensitive
to such small velocity shifts (George \etal\ 1997).  In any case,
velocity shifts might be most difficult to observe in type-2 Seyferts
because the wind velocity vector could be almost perpendicular to our
line-of-sight.

\subsection{Constraints from emission}

The strength of the iron K$\alpha$ lines in the group C 
sources is a strong indication that they arise via reprocessing.  As
stated above, the majority of the line emission occurs close to
6.4~keV, suggesting an origin via reflection from the 
absorbing material. This component of the line
profile is expected to be free of the relativistic
effects evident in the iron K$\alpha$ line profiles of Seyfert~1
galaxies (N97). Fig.~4 shows the individual ratios in the 
iron K$\alpha$ regime, from 
reprocessed sources with the most significant iron 
lines, i.e. NGC~1068, Mrk~3, NGC~2992 and NGC~4945. Ratios 
are shown for the summed SIS plus GIS data, and the SIS data alone.  
The SIS data alone have 
superior energy resolution, but unfortunately our systematic method of
analysis has left few significant SIS data bins at high energies.  We
note that Iwasawa \etal\ (1996) obtained a clearer SIS line profile 
for NGC~1068 by optimizing the data screening for that observation
(the conservative criteria used in N97 and Paper~I yield a low SIS
exposure compared to the on-time of the observation in this case).

The SIS data can provide some useful information when the data/model
ratios are co-added.  The energy-scale of each SIS spectrum was
redshift-corrected to the rest-frame of the source and then the
average SIS instrument ratio was calculated utilizing 
all of the group C sources (versus the best continuum 
model in each case, from Paper~I).  Fig.~5 shows the result of this
summation. The average SIS data/model ratio is shown, and the dotted
Gaussian profile represents the SIS instrument response for a narrow
line observed at a rest-energy of 6.4 keV.  This plot does not show the 
actual line profile, rather the counts ratio compared to the best continuum 
model (to get the
true profile you have to multiply the ratio plot by the continuum form). 
Excess flux is evident 
blueward and redward of 6.4 keV.  The former seems most likely to
indicate the presence of high-ionization species of iron K$\alpha$,
which are present in NGC~1068, Mrk~3, NGC~4945, and NGC~2992 when
tested on an individual basis (Paper~I).  Mrk~463E and Mrk~273 did not
have sufficient signal-to-noise to detect features of a similar
relative strength. All of the sources showed a significantly broad
line when the iron K$\alpha$ regime was modelled with a single
gaussian, so the sources either have multiple unresolved components, 
like NGC~1068 (Marshall \etal\ 1993), or the iron K$\alpha$ lines 
are broadened. 

It therefore seems most likely that the spectra contain contributions
both from Compton reflection and the scattering gas, although as noted
above, if the latter has multiple ionization states and the dominant
one is relatively cool, this could account for a strong 6.4 keV
component without the need for reflection. Of particular interest in
this regard is the observation of a red wing to the line. 
This signature -- which amounts to $\sim 10$~per 
cent of the core flux -- is not expected to be produced 
by scattering from an optically 
thin medium.  The same feature is observed with the same equivalent width 
in the iron K$\alpha$ profile of Seyfert~1 galaxies (N97). 
Iwasawa \etal\ (1996) noted the presence of a red-wing
in the iron K$\alpha$ line profile of NGC~1068. Those authors  
noted that it was consistent with the 
``Compton shoulder'' expected from downscattering of 6.4 keV photons 
in a reflecting medium. 
The Mrk~3 and NGC~4945 profiles appear to show the 
same effect but NGC~2992 does not (Fig.~4,  although the data do not
allow us to place interesting constraints in this respect,  Mrk~463E
and Mrk~273 are inconclusive due to a lack of signal-to-noise in the
5-7 keV regime.)  Iwasawa \etal\ (1996) state that the red wing in
NGC~1068 is more likely to be attributed to a Compton shoulder than to
the relativistic effects observed in Seyfert~1 galaxies, as the
contribution of the relativistic disk should be observed only in
scattered light. However, if the continuum is also scattered, the
relativistic disk component would have the same equivalent width with
respect to that continuum as it does when it is directly
observed. This would result in a red wing similar to that observed. 
The fact that some of the line
profiles extend to energies lower than could easily be produced by
Compton scattering in a cold medium suggests that the red wings may
indeed be associated with the relativistic disk component seen in
scattered light. However, uncertainties in the continuum modelling 
used to produce the line profiles may make such a conclusion premature. 

The presence of the high-ionization lines make the average line
profile from group C significantly different to that obtained for a sample of
Seyfert~1 galaxies (N97) (which shows a strong broad red-wing but no
high-ionization species of iron K$\alpha$).  While the two types of
Seyfert galaxy may well contain iron K$\alpha$ line components from
all the same regions the high-ionization lines would be swamped by the
direct continuum in the Seyfert~1 case.

If these high-ionization iron K-shell lines arise in the scattering gas,
then in the model of Heisler, Lumsden and Bailey (1997), we might
expect to preferentially detect these species in sources for which a
hidden BLR has been detected in the scattered optical light. This is
because their model places the scattering region within the obscuring
torus, so hidden BLRs are visible when sources are observed from
lines-of-sight which skim the edge of the torus, i.e. for a narrow
range of observing angles. For sources viewed at a favorable angle 
the torus hides the central engine and BLR, but allows direct
observation of the scattering gas. For sources viewed more edge-on,
the scattering region is also hidden. Our data are consistent with
that idea, in that we detect these line species in all the 
sources where we see the BLR in scattered  optical light and 
for which we have sufficient signal-to-noise ratio in the X-ray 
spectra to tell.

We would expect to observe emission lines other than iron-K from the
scattering gas, particularly at soft X-ray energies, and indeed many
such lines are observed (Paper I). Ideally, then, one would wish to
compare detailed photoionization models for the scattering region with
our data to determine the physical characteristics of the gas.
However, some ambiguity remains as to whether the observed soft X--ray
lines arise via nuclear processes, or are related to starburst
activity. Furthermore, high-ionization iron--K species are detected in
sources such as NGC~5135,  which has a strong starburst component. 
This raises the question as to whether some of the iron K$\alpha$ emission
might arise from non-nuclear processes.  Clearly, separating the emission
of the Seyfert from that of the starburst is necessary for a conclusive
answer.

\section{Starburst contributions to the X-ray Flux}

Starburst emission is a well-documented phenomenon in many Seyfert 2
galaxies. It is not yet clearly established as to 
whether starburst activity occurs at the same
level in Seyfert 1 galaxies. Some studies indicate 
a similar star formation rate in Seyfert~1 galaxies and 
starburst galaxies (Yamada 1994) while others 
indicate that the host 
galaxies of type-2 nuclei have significantly higher levels of
starburst activity than type-1 nuclei (Maiolino \etal\ 1995). 
If it does occur at the same level in the two types of 
Seyfert, the X-ray emission
associated with the starburst would be swamped by the Seyfert 1
nuclei.  For a sample of obscured AGN, we would expect some observable
differences as the degree of nuclear obscuration varies between
sources and hence the contrast is changed between contributions from the
host galaxy versus the nucleus.  In Paper I we found that in the 0.5-4.5
keV band the estimated average contribution from starbursts in the
host galaxy was $\sim 60$\% for the Seyfert~2 galaxies, versus 2\% for
the NELGs.  We also noted that soft X-ray lines were observed in many
Seyfert~2 galaxies, but in none of the NELGs (Paper~I).  As strong
soft X-ray lines are seen in some starburst galaxies (Ptak \etal\
1997), these facts necessitate careful consideration of the the
starburst contribution to the X-ray flux.

Unfortunately, the \asca\ data do not allow us to spatially deconvolve
the X-ray lines produced by starburst activity from those produced in
the scattering gas. However, Wilson \etal\ (1992) find that the
starburst disk in NGC~1068 can provide a 2-10 keV flux of 
$\sim 2-8 \times 10^{-12} {\rm erg\ cm}^{-1} {\rm s}^{-1}$, consistent 
with all 
of the {\it ASCA} flux.  So, although the high equivalent widths and
energies of the iron K$\alpha$ lines link them with reflection or
scattering of nuclear radiation, some or all of the other X-ray
emission lines may be attributable to starburst activity.

Examining the level of starburst ``contamination'' for targets in 
group C, we find just one source which is not 
significantly contaminated by a starburst component, Mrk~3 (Paper~I
and Pogge \& De Robertis 1993). This source is particularly
interesting as, despite the lack of starburst emission, it does show
significant soft X-ray emission lines. Mrk~3 is  one of the brightest
Seyfert~2 galaxies with a hidden BLR.  Detailed analysis of the
optical data has shown the broad lines have a polarization of 20\%
(Tran 1995b). Although there has been some debate
over the Hubble type of the host galaxy for Mrk~3, it is now thought
to reside in an elliptical host galaxy (Jenkins 1981, Wagner 1987),
and it has a radio jet with knots and a large bright radio lobe
(Kukula \etal\ 1993). 
This source offers us the best opportunity available to
date to obtain information about conditions in the X-ray reprocessing
regions of a Seyfert~2 galaxy.

\section{A case study: Mrk~3}

We constructed a spectral model table of the emission spectrum from
a slab of photoionized gas illuminated by an ionizing continuum 
using the ION photoionization code. ION includes the important 
excitation and ionization processes, full temperature and 
radiative transfer solutions, while emission, absorption 
and reflection by the gas are calculated asuming thermal 
and ionization equilibrium. The model tables we use assume 
a photon index 
$\Gamma=2.0$ in the 0.2 -- 50 keV range, and 
$\Gamma=1.5$ from 1.6 -- 40.8 eV. The optical to 
X-ray energy index is assumed to be $\alpha=1.5$.  
Cosmic abundances were used, and a constant density 
$n_H=10^{10} {\rm cm}^{-3}$ throughout the slab. The X-ray 
ionization parameter is defined as 
\begin{equation}
\label{eqn:U_X}
U_X = \int^{10\ {\rm keV}}_{0.1\ {\rm keV}}
\frac{Q(E)}{4 \pi r^2 n_{H} c} dE
\end{equation}
where $Q(E)$ is the number of photons at energy $E$,
$r$ the distance from the source to the illuminated gas.
For more details of the ION code see Netzer (1993; 1996). The 
absorption of 
this component was fixed at the Galactic value $8.7 \times 10^{20}
{\rm cm}^{-2}$.  In addition to this component we allowed a pure
Compton-reflection spectrum utilizing the model of Magdziarz \&
Zdziarski (1995) plus a gaussian line (since their reflection model does not 
include line emission).  We assumed reflection
of a $\Gamma=2.0$ continuum (with no exponential cut-off; the
data do not allow us to determine whether any cut-off is present) from
relatively neutral material (only H and He are ionized) viewed
face-on.  This model provided a good fit to the data
yielding $\chi^2=349/328$. 

\subsection{The Hard X-ray Spectrum}

The energy at which the primary continuum becomes 
visible depends on the column obscuring it. 
If the nucleus of Mrk~3 was obscured by a 
column $\sim 10^{24} {\rm cm}^{-2}$ as observed in NGC~4945, then 
the direct continuum component should become observable close to 
$\sim 10$ keV. In fact the {\it ASCA} spectrum does show a small 
hard tail compared to the Compton-reflection model noted above. 
The {\it ASCA} fit improves by  $\Delta \chi^2 \sim$11 on addition 
of a highly absorbed  power-law component of fixed index $\Gamma=2$.  
This component is absorbed by a column of 
$N_{H}=1.3^{+2.2}_{-0.6} \times 10^{24} {\rm cm}^{-2}$, and 
provides 25\% of the 5-10 keV observed flux.  We estimate 
the intrinsic (unabsorbed) 2-10 keV luminosity of this nuclear component 
to be $\sim 10^{43} {\rm erg\ s}^{-1}$. 
The reflecting medium is implied to subtend $\Omega \sim 2\pm 1\pi$ 
steradians to the intrinsic continuum source.
This model is shown in Fig.~6, along with the data/model ratio.

Iwasawa \etal\ (1994) find a different solution to the 
{\it ASCA} spectral data, with  a model in which 
the direct absorbed continuum component dominates the 3 -- 10 keV spectrum, 
absorbed by $N_{H} \sim 4 \times 10^{23} {\rm cm}^{-2}$. 
No reflection component is included in their model. 
However, the observed equivalent width in iron K$\alpha$ 
is $977^{+193}_{-137}$ eV (when modelled as a narrow line, Paper~I), 
which requires a 
column of $10^{24} {\rm cm}^{-2}$  in order for the line 
to be produced by transmission (Fig.~1), assuming a 
Leahy \& Creighton (1993) geometry. The upper limit on column 
yielded by the Iwasawa \etal\ (1994) solution is 
$5.8 \times 10^{23} {\rm cm}^{-2}$, which is too low to produce 
the line, assuming the Leahy \& Creighton geometry is 
applicable.

Our model, which includes the Compton-reflection component, 
yields a better fit by $\Delta \chi^2 =59$ (obtained by 
comparing the fit of their model and ours using our datasets).  
The inclusion of 
a Compton-reflection component also provides a natural explanation for the 
very flat $\Gamma=1.30\pm0.3$ spectrum observed in the {\it Ginga} data 
(Awaki \etal\ 1990). The inclusion of the reflection component is the reason 
we obtain a higher column for the direct component than Iwasawa \etal\ (1994), 
in fact our column of $\sim 10^{24} {\rm cm}^{-2}$ is consistent with the 
strength of the iron K$\alpha$ line. That line 
is now consistent either with an origin in the line-of-sight material 
or via Compton reflection alone. The indication that the line flux 
variations are correlated with the hard X-ray continuum variations 
favors the latter (see \S6.3). 

\subsection{The soft X-ray Spectrum}

The soft part of the spectrum (below $\sim 3$ keV) is well 
modelled by the emission from a photoionized plasma
with $U_{X}=4.8^{+1.5}_{-1.3}$ and $N_{H}=3.5^{+1.7}_{-1.4} \times
10^{23} {\rm cm}^{-2}$. This highly ionized gas can produce the 6.96 keV
component of the iron line and the Si and S lines and scatters some of
the nuclear radiation into our line-of-sight.  This component is much
more highly ionized than the zone of material comprising the warm
absorber evident in {\it ASCA} observations of Seyfert~1 galaxies
(when a single-zone model is assumed for the latter).  
The temperature of the gas is $\sim 10^5$ K, consistent with that 
of the optical scattering gas  in NGC~1068 (Miller, Goodrich \& Mathews 
1991)

However, this soft X-ray flux might have some contribution 
from shocked gas. Recent {\it HST} observations (Capetti \etal\ 1995) have 
resolved a sub-arcsec region of continuum emission, which they propose could 
either be associated with gas which is shocked by passage of the jet, or 
the BLR scattering gas. Those authors also show a $\sim 2 \arcsec$ 
S-shaped region of narrow line gas, although this gas is 
photoionized by the central AGN, it is too cool to produce the 
soft X-rays emission observed. 

The absorption-corrected 2-10 keV luminosity in the ionized-gas component 
alone  is $5 \times 10^{41} {\rm erg\ s}^{-1}$. 
The scattered fraction of flux is proportional to the 
product of the scattering efficiency and the solid angle 
of the scattering material. Assuming the scattering gas dominates 
over shock-heated gas, then our X-ray data indicate a 
scattered X-ray fraction of  $\sim 5\pm 1$\%. This is 
comparable to the few percent generally derived from optical 
measurements of type-2 AGN (e.g. Pier \etal\ 1994).  
The estimate of intrinsic 2-10 keV luminosity from 
the \verb+[+O{\sc iii}\verb+]+ line flux  
is $2 \times 10^{44} {\rm erg\ s}^{-1}$, yielding an 
estimate of 0.25\% for the scattered fraction. 
Alternatively, using the Mulchaey (1994) relationship between
infrared and hard X-ray luminosity 
we estimate an intrinsic luminosity of 
$6.6 \times 10^{43} {\rm erg\ s}^{-1}$ and a scattered fraction 0.75\%.

\subsection{Constraints from variability}

If the hard X-ray emission is transmitted through the obscuring
material in our line-of-sight, then we will observe the hard X-ray
variations of the nucleus directly.  Alternatively, if the hard X-ray
emission is dominated by a reflected component, then the light curve
of the reprocessed emission will lag the primary emission and be
smoothed. The degree of lag and the smoothing depend on the location
and size of the reprocessor and its geometry.

The soft X-ray flux has not varied over the
$\sim 13 $ year baseline for which we have flux measurements (Iwasawa
\etal\ 1994), suggesting that it comes from an extended region. 
If this flux is associated with the $0.35 \times 1 \arcsec$ bar 
of continuum emission resolved by {\it HST} (Capetti \etal\ 1995), 
then it originates in a region several hundred pc in size. 

However, the hard X-ray emission of Mrk 3 is variable.  The fastest
historical change in the hard flux was a factor of 2 drop between the
{\it Ginga} and {\it BBXRT} observations, taken one year apart. The 6.4 keV
line flux appears to vary in a way which is consistent with that
change in the continuum  while the soft X-ray flux remains steady, 
ruling out a lag between the reprocessed and 
nuclear flux as the explanation for the high equivalent width.

The observed variability places a rough constraint on the size of the
variable source to be $< 0.3$ pc.  The shape of the spectrum suggests
that the nucleus is not observed directly and therefore this
constraint applies to the reprocessing material, rather than the
central source.  This could have a small scale-height, in fact
Gallimore \etal\ (1996) suggest that a pc-scale maser disk with
toroidal geometry exists in NGC~1068, based on a water maser
observation. This disk is observed at high inclination ($82^{\rm o}$)
and can provide sufficient opacity to block the X-ray continuum below
10 keV.

\section{Discussion and Conclusions}

Consideration of the iron K$\alpha$, \verb+[+O{\sc iii}\verb+]+ line and 
X-ray variability indicates that the {\it ASCA}
observations are dominated by reprocessed X-rays in the cases of
NGC~1068, NGC~4945, NGC~2992, Mrk~3, Mrk~463E and Mrk~273.  

The iron K$\alpha$ line profiles of these sources show significant 
contributions from neutral and high-ionization species of iron.  
Thus it appears that Compton reflection, the hot scattering gas 
and/or starburst emission contribute to the 5-7 keV regime.

Mrk~3 shows negligible starburst contamination, so the X-ray spectrum
is probably dominated by processes intimately related to the active
nucleus.  The {\it ASCA} spectrum can be
described by hot scattering gas in the soft X-ray regime, with $U_{X}
\sim 5$, $N_{H} \sim 4 \times 10^{23} {\rm cm}^{-2}$.  This 
material is more highly ionized than the zone of material comprising
the warm absorber seen in Seyfert~1 galaxies, but may 
have some contribution from shock-heated gas associated with the jet.  
Estimates of the scattering fraction 
cover 0.25 -- 5\%. The hard X-ray spectrum of 
this source appears to be dominated by a Compton-reflection component 
although there is evidence that the primary continuum
component becomes visible close to $\sim 10$ keV.  

It is obvious that a big problem with analysis of X-ray observations
of type-2 Seyferts to date has been the inability to deconvolve the
starburst emission (and any other extended thermal gas) from that in
the immediate environment of the nucleus (and while {\it ROSAT}
provided several arcsecond spatial resolution it offered data over a
narrow bandpass with low spectral resolution).  {\it HST} 
observations have now resolved a 1-2 arcsec region in which the UV
scattering occurs in NGC~1068 (Antonucci \etal\ 1994), corresponding
to a region $\sim 110-220$ pc in size.  The {\it AXAF} High Resolution
Camera (HRC) will allow sub-arcsecond X-ray imaging. Assuming the UV
scattering region is closely related to the X-ray scattering region,
then we will be able to spatially resolve the X-ray ``mirror'' for the
first time in NGC~1068, other nearby sources such
as NGC~4945 and NGC~1808 and we will be able to resolve the 
bar of emission observed in Mrk~3 by {\it HST}. If the 
X-ray scattering gas turns out to be 
the same gas which is observed as a warm absorber in Seyfert~1
galaxies, then such imaging may map the distribution of the warm
absorber. As the starburst emission
often extends over regions many tens of arcseconds or larger for
nearby AGN, AXAF spectra should allow us to separate the starburst and
nuclear components for at least some AGN.  {\it Beppo SAX} and {\it XTE} also
offer an opportunity to acquire X-ray spectra above 10 keV, which might
help us determine the true nature of Mrk~3.

\section{Acknowledgements}
We are grateful to \asca\ team for their operation of the satellite, 
to Keith Gendreau for discussions on \asca\ calibration issues, 
Hagai Netzer for use of model tables generated using the 
photoionization code ION and Andy Ptak for comments on the properties of 
starburst galaxies. This research has made use of the Simbad database, 
operated at CDS, Strasbourg, France; of the NASA/IPAC Extragalactic database,
which is operated by the Jet Propulsion Laboratory, Caltech, under
contract with NASA; and data obtained through the HEASARC on-line
service, provided by NASA/GSFC. We acknowledge the financial support
of Universities Space Research Association (IMG, TJT) and 
the National Research Council (KN).

\newpage

{\bf Figure Captions}

Fig 1 - Equivalent width of the narrow 6.4 keV line 
versus absorbing column (from Paper~I). The dot-dash line shows 
the line equivalent 
width expected from a uniform shell of material encompassing the
continuum source (Leahy \& Creighton 1993). The dashed line  shows the
predicted equivalent width from reflection as a function of \nh,
assuming that only the power-law is absorbed, but that the reflection
component remains unchanged. Data points are annotated with an
abbreviation of the source name. Group A sources are marked as circles, 
group B sources as stars and group C sources as squares (see \S3.3). 
A few of the lowest signal-to-noise datasets have not been classified. 

Fig 2 - The predicted intrinsic 2-10 keV luminosity, based upon 
the \verb+[+O{\sc iii}\verb+]+ line flux (see text for details) versus the 
observed (absorption-corrected) luminosity. We have annotated those 
 sources having predicted luminosities more than a factor of 15 
greater than the observed values. Again, group A sources 
are marked as circles, group B sources as 
stars and group C sources as squares (\S3.3). 

Fig 3 - The emission spectrum from a warm absorber including 
scattered continuum emission (from a $\Gamma=2$ power-law) plus 
thermal and line emission from the hot gas. The gas has
an ionization parameter $U_X \sim 0.12$ and column
$N_H \sim 4 \times 10^{22} {\rm cm}^{-2}$, based on
the analysis of {\it ASCA} observations of Seyfert~1
galaxies (see George \etal\ 1997). 

Fig 4 - The data/model ratios versus the best-fit simple 
continuum model for NGC~1068, Mrk~3, NGC~2992, and NGC~4945 (from Paper~I). 
The X-axis shows the observed-frame energy. In the left hand panels 
data from all four instruments have been combined 
for clarity. In the right hand panels, the two SIS datasets 
have been combined. The data are in 50 eV bins up to 3 keV, 
100 eV bins between 3 and 7 keV and 200 eV bins in the 7-10 keV range. 
An excess of counts is generally evident in the 5-7 keV data,
versus the continuum model. This indicates the presence of strong iron
K-shell emission lines in most objects.

Fig 5 - The average data/model ratio in the iron K$\alpha$ regime. The 
group C datasets were used to create the mean ratio. Each dataset was 
corrected to the rest-frame energy before the objects were combined 
and so the  X-axis shows the rest-frame energy. 
The dotted Gaussian profile represents the SIS instrument 
response for a narrow line observed at a rest-energy of 6.4 keV. 

Fig 6 - The best fit model to the {\it ASCA} 
spectral data for Mrk~3, along with the data/model ratio. 
The SIS and GIS ratios have been combined for clarity of illustration. 
The model comprises an emission spectrum from highly 
ionized gas, a Compton-reflection component and an absorbed power-law 
(see \S6). A Galactic column of $8.7 \times 10^{20} {\rm cm}^{-2}$ 
covers all components. 







\clearpage
\begin{deluxetable}{l l c c c c c}
\tablecaption{The \asca\ Seyfert 2 sample. \label{tab:sample}}
\tablehead{
\colhead{Name} &  \colhead{{RA}$^{a}$}
&  \colhead{{DEC}$^{a}$} & 
\colhead{z$^a$} & \colhead{Class$^a$} & \colhead{$L_X$} 
& \colhead{$L_{\verb+[+O\protect\small III\verb+]+ }$}
}
\startdata
MCG-01-01-043 & 00 10 03.5 & -04 42 18  &0.0300 & S2 & 42.88 & \nodata \nl
NGC 526A    & 01 23 55.1 &-35 04 04 &  0.0192  &NELG$^b$ & 43.77 & 43.49 \nl
{\bf NGC 1068}& 02 42 40.8 & -00 00 47   & 0.0038&S2$^b$& 41.15 & 43.95  \nl
NGC 1667  & 04 48 37.2  &-06 19 12  & 0.0152 & S2 & 40.46 & 42.69 \nl
E0449-184 &04 51 38.8 &  -18 18 55 & 0.338 & S2 & 44.86 & \nodata \nl
NGC 1808    & 05 07 42.3 & -37 30 46 & 0.0033 & SB/S2 & 40.62 & 40.69  \nl
NGC 2110    &05 52 11.4 & -07 27 22  & 0.0076& NELG &  42.82 & 42.49 \nl
{\bf Mrk 3}& 06 15 36.3 &71 02 15  & 0.0135& S2$^b$& 42.33 & 44.30 \nl
{\bf NGC 2992}& 09 45 41.9  & -14 19 35 &0.0077& NELG$^b$&42.05 & 43.23 \nl
MCG-5-23-16 &09 47 40.2 &-30 56 54 &0.0083&NELG$^b$& 43.43 & 42.69  \nl
NGC 4507 &  12 35 36.5 &  -39 54 31& 0.0118 & S2 & 42.28 & 43.68  \nl
{\bf NGC 4945}&13 05 26.2 &-49 28 16&0.0019 & S2 & 40.53 & \nodata  \nl
NGC 4968  &13 07 06 & -23 40 43 & 0.0100 & S2 & 42.45 & 42.50  \nl
NGC 5135   &13 25 44 &-29 50 01 &0.0137 & S2 & 43.16 & 43.11 \nl
NGC 5252  & 13 38 15.9 & 04 32 33  & 0.0230 & S1.9 & 43.24 & 43.56 \nl
{\bf Mrk 273}&13 44 42.1 & 55 53 13 &0.0378 & S2 & 42.45 & 44.03 \nl
{\bf Mrk 463E}& 13 56 02.9 &18 22 19 &0.0500 & S2$^b$& 42.75 & 44.75 \nl
NGC 5695    &14 37 22.0 &  36 34 04& 0.0141 &S2 & $<40$ & 42.65 \nl
NGC 6251    &16 32 31.9 &82 32 17 &0.0230 & S2 & 42.50 & 41.94  \nl
NGC 6240-49 & 16 52 59.3 &02 23 59 & 0.0245&S2& 43.61 & 42.69 \nl
ESO 103-G35 &18 38 20.2 &-65 25 42 & 0.0133 & S2/NELG& 43.14 & 42.84  \nl
IC 5063&20 52 02.9 & -57 04 14  & 0.0113 & S2$^c$ & 43.15 & 43.57  \nl
NGC 7172 &22 02 02.1 & -31 52 12  &0.0086&S2/NELG$^b$& 43.22 & \nodata \nl
NGC 7314 & 22 35 45.7 &-26 03 03 &0.0047 & S1.9/NELG & 42.59 & 41.63 \nl
NGC 7582 & 23 18 23.2  & -42 22 11  & 0.0053 & S2/NELG$^d$&42.32 & 42.65 \nl
\tablenotetext{a}{From the NASA Extragalactic Database (NED); \nl 
$^b\ ^c\ ^d$ Polarized broad lines have been detected: (b) Tran 1995, 
(c) Inglis \etal\ 1993, (d) Heisler, Lumstron, Bailey 1997}
\tablecomments{Group C sources are bold-face. Column 1: Name; 
Columns 2 \& 3: RA \& DEC,  J2000; Column 4: Redshift; Column 5: 
Seyfert type, SB= starburst galaxy, NELG=Narrow-Emission-Line Galaxy; 
Column 6: Observed 2-10 keV luminosity (absorption corrected, but see \S4.1); 
Column 7: Intrinsic 2-10 keV luminosity inferred from \verb+[+O{\sc iii}\verb+]+  fluxes}
\enddata
\end{deluxetable}

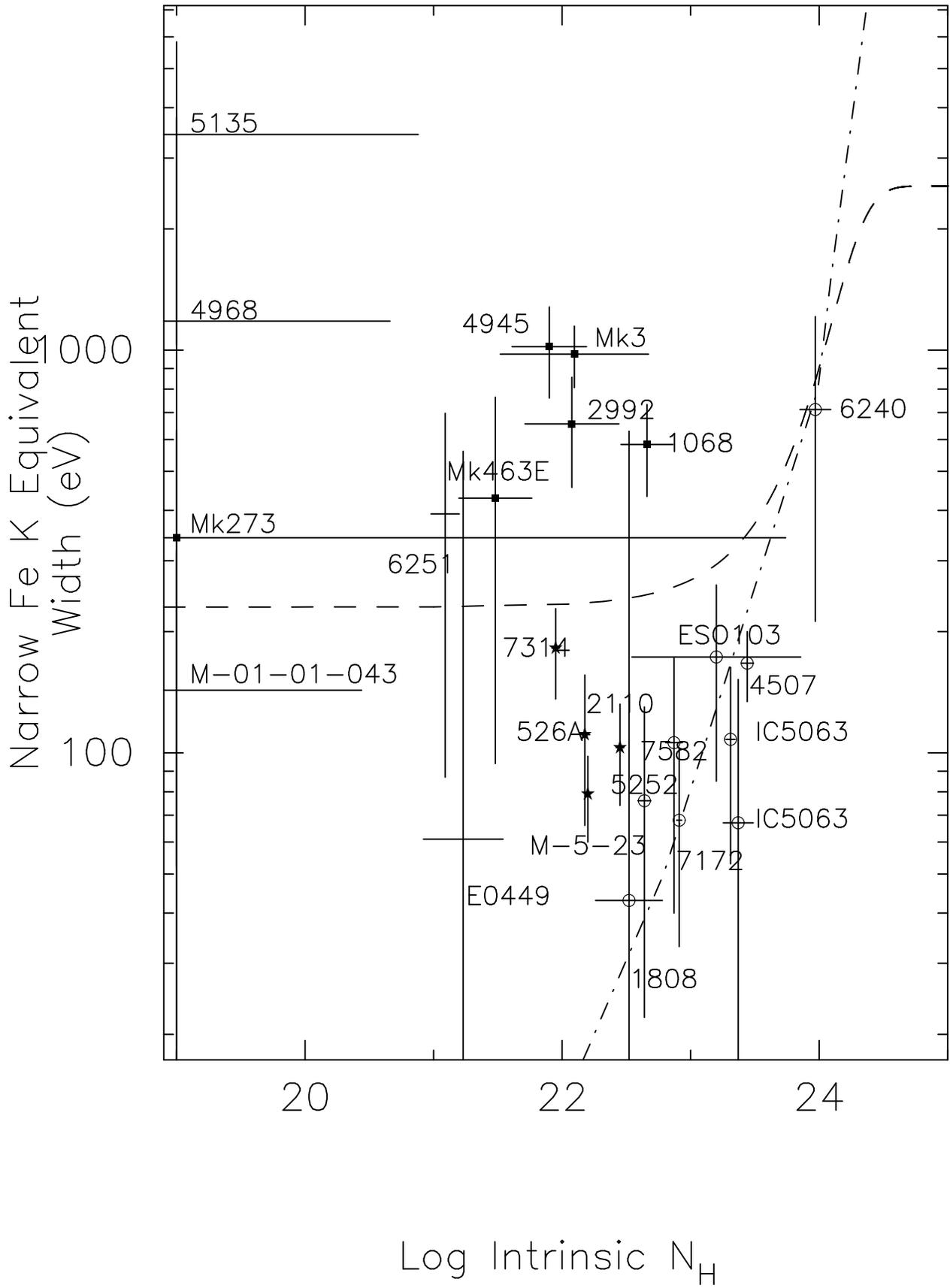
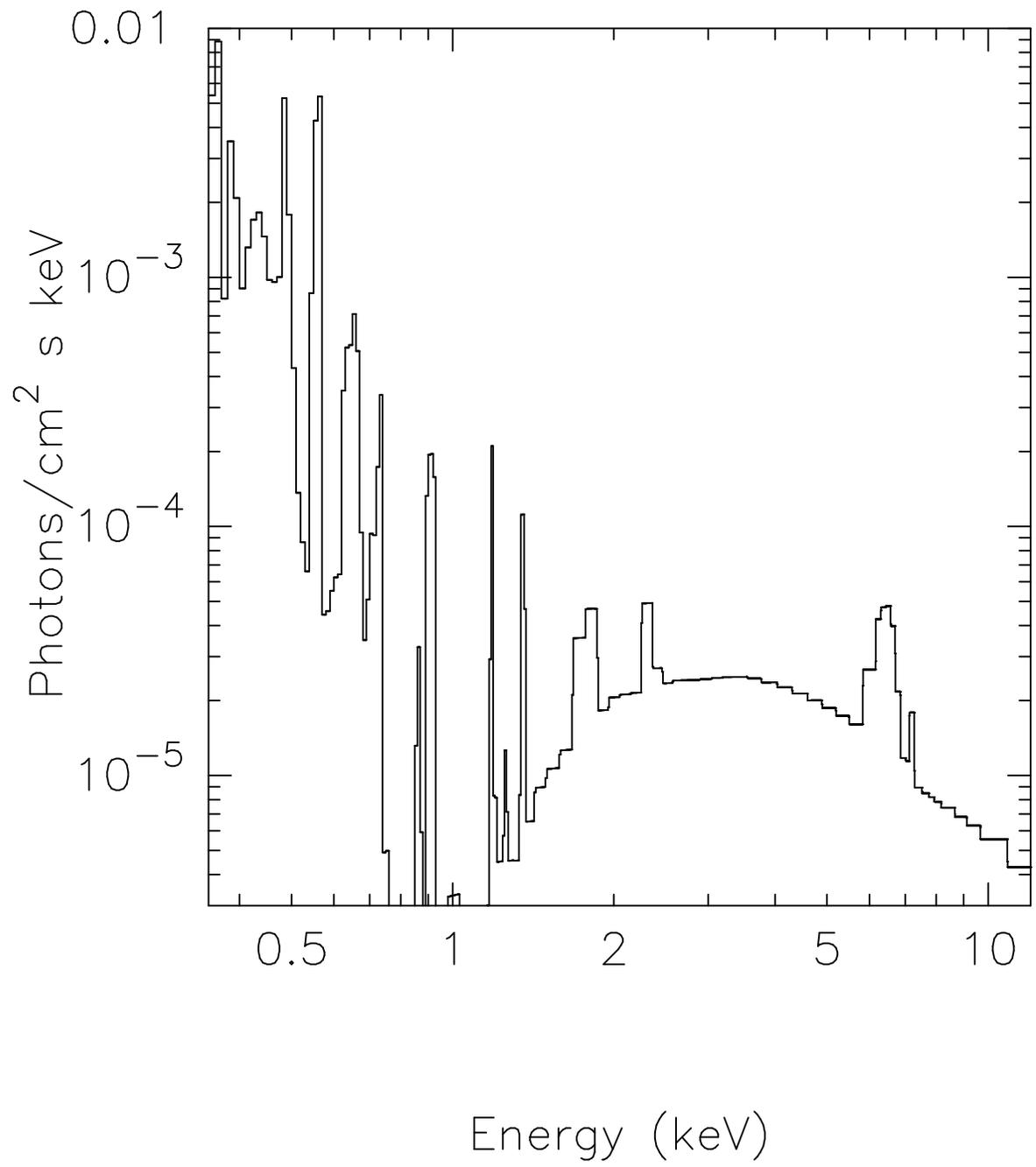
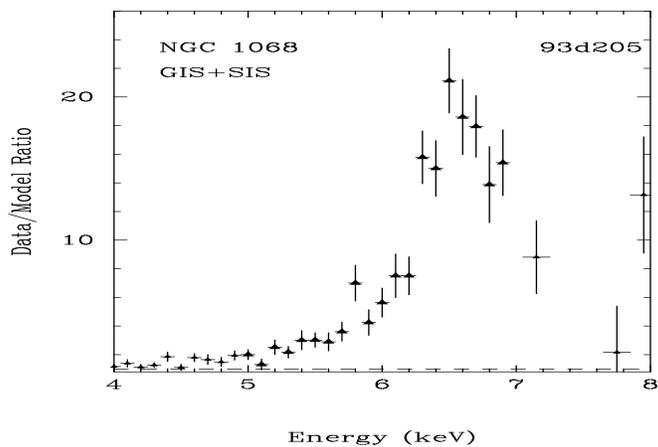
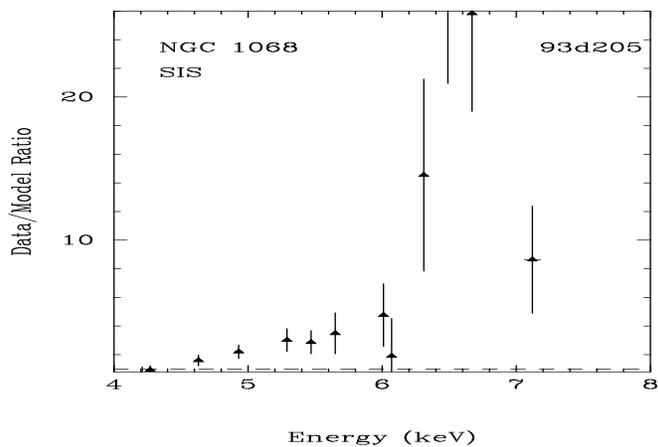
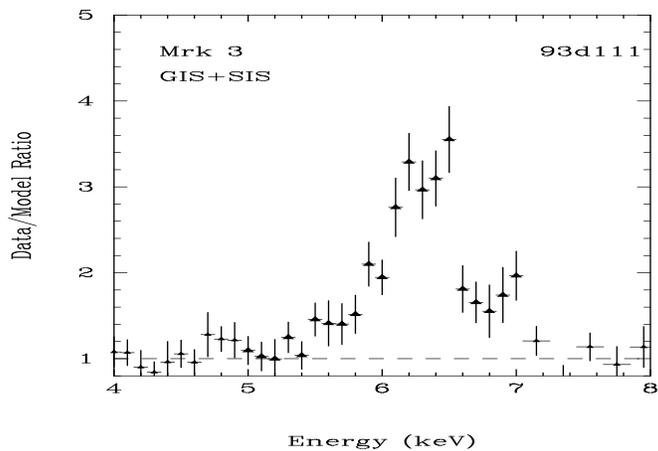
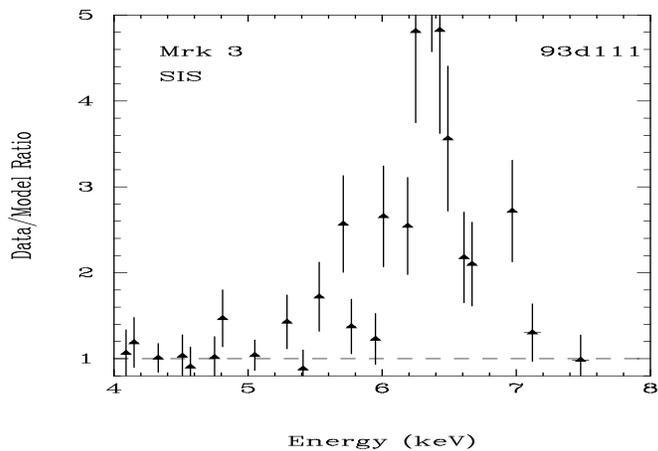
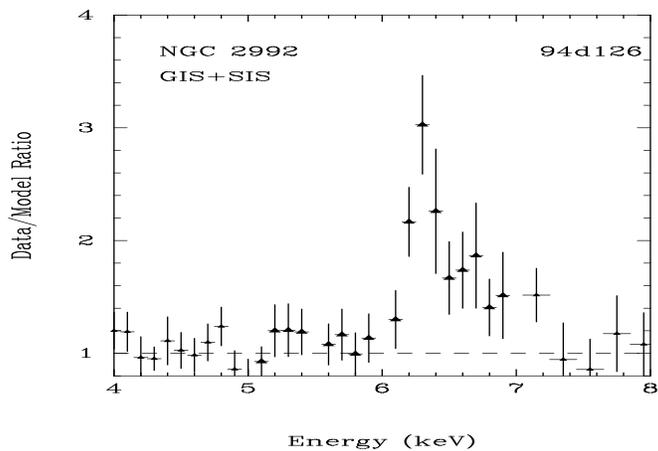
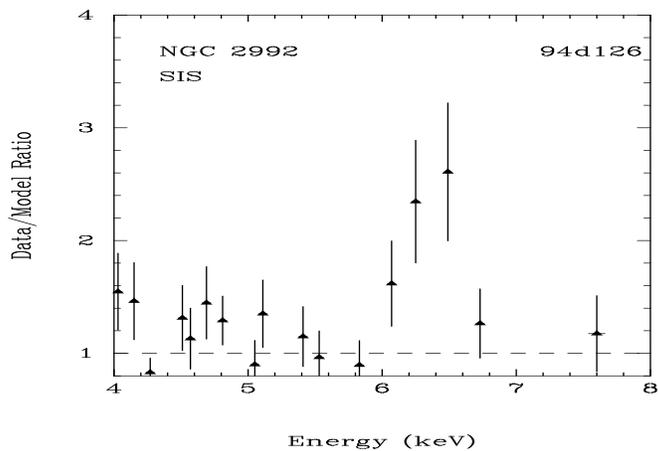
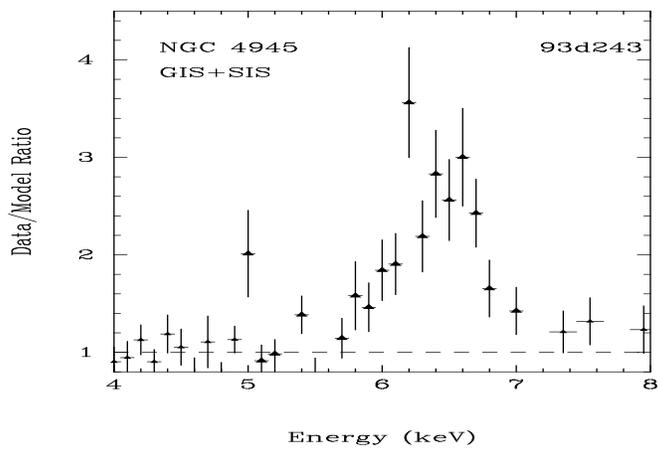
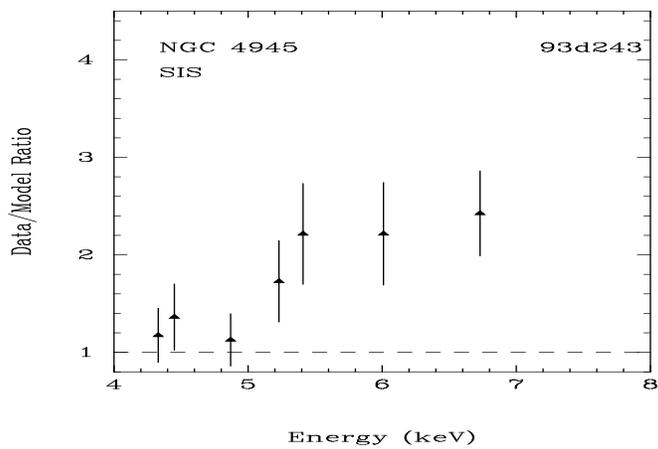
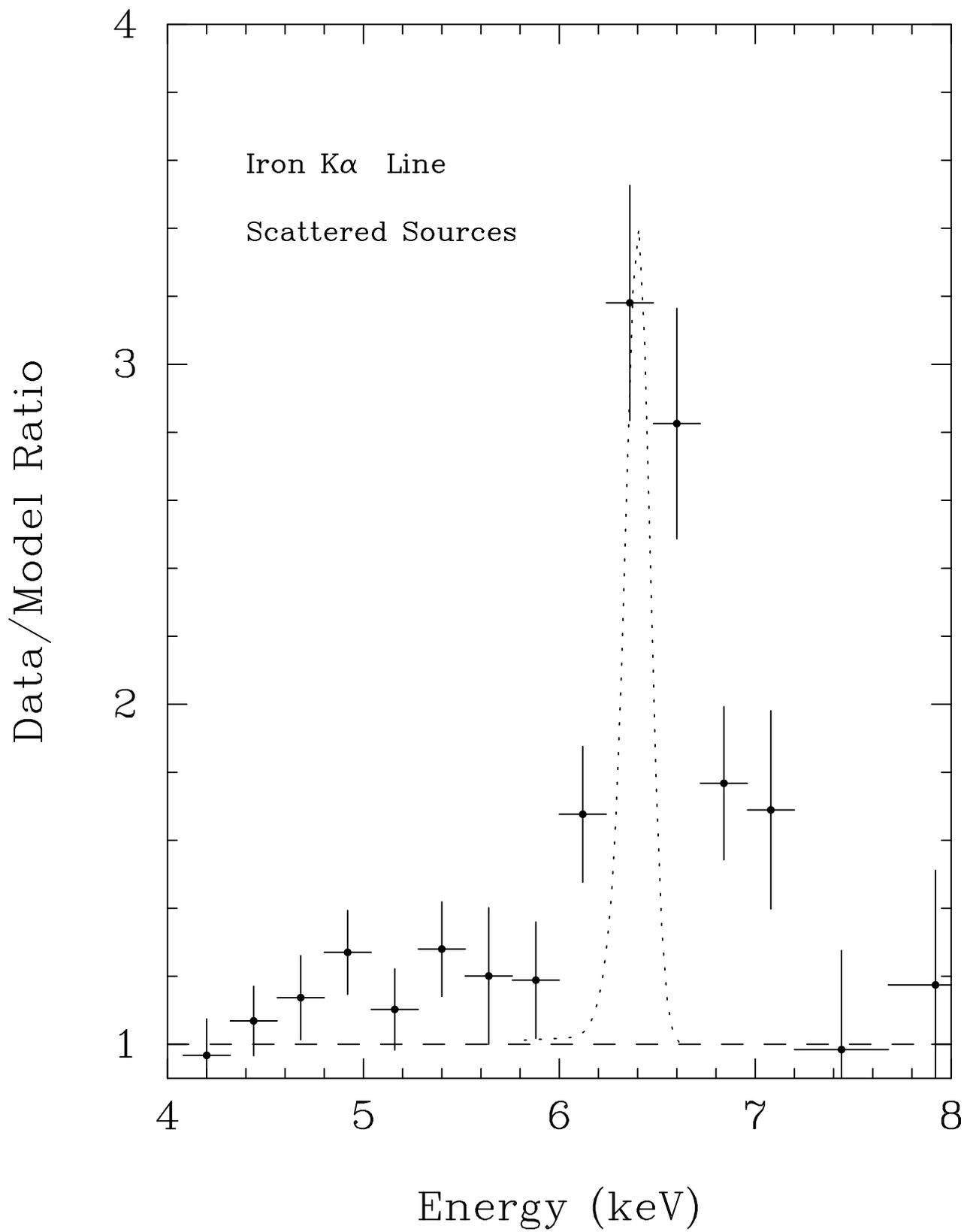
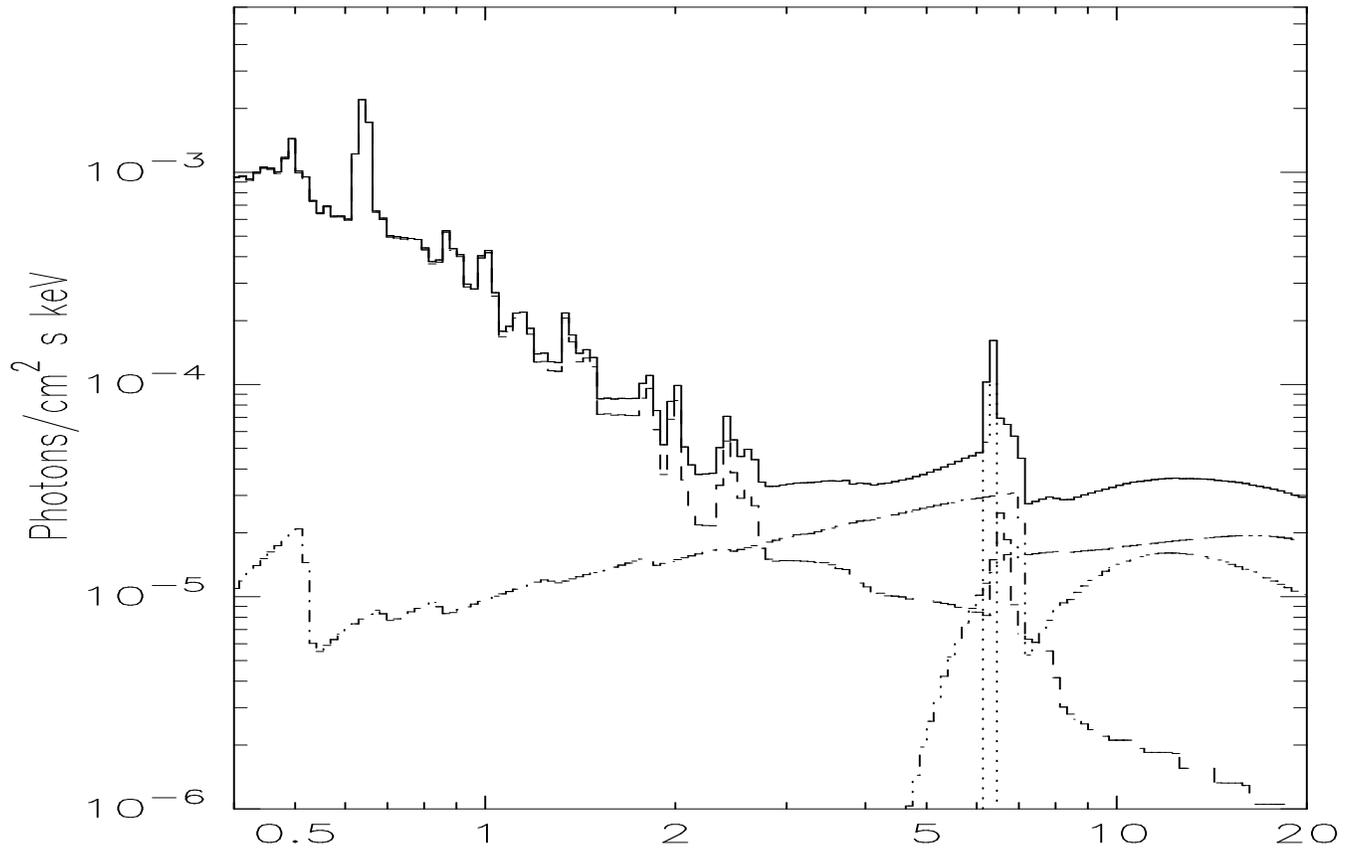
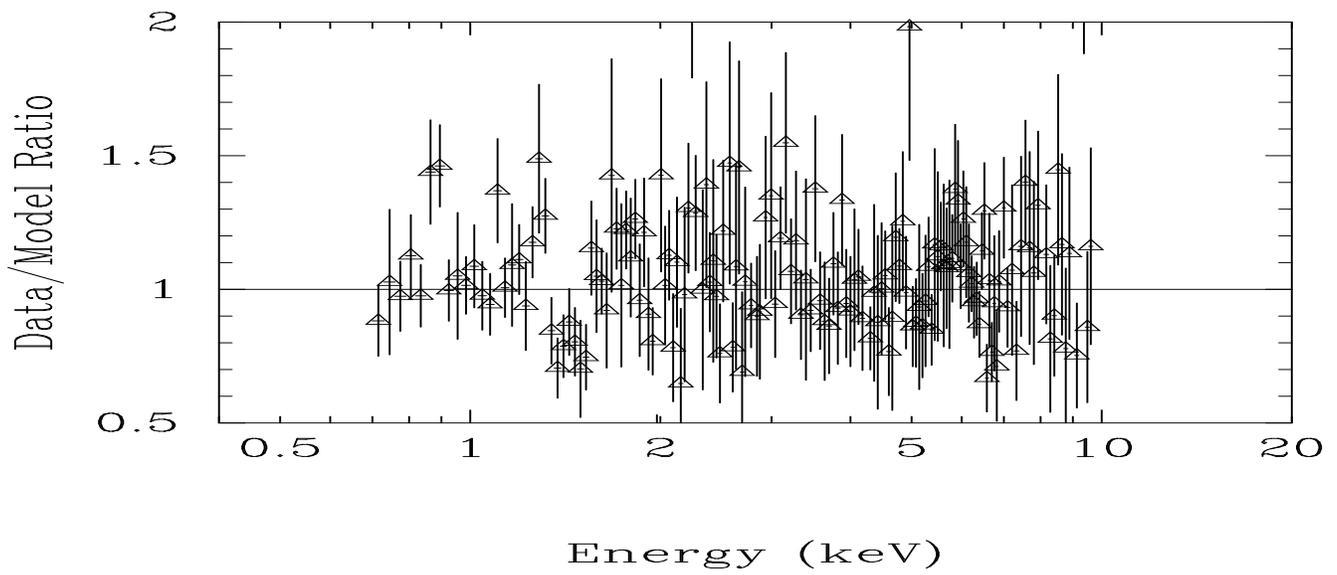
\end{document}